\documentclass[epj]{svjour}
\usepackage{graphicx}
\usepackage[usenames,dvipsnames]{xcolor}

\begin{document}
\newcommand{\edit}[1]{\textcolor{red}{#1}}

\title{Surface microswimmers, harnessing the interface to self-propel}
\author{G. Grosjean, M. Hubert, Y. Collard, S. Pillitteri \and N. Vandewalle
}                     
\authorrunning{G. Grosjean \emph{et al.}}
\institute{GRASP, Universit\'e de Li\`ege, All\'ee du 6 Ao\^ut 19, 4000 Li\`ege}
\date{Received: date / Revised version: date}
%
\abstract{
In the study of microscopic flows, self-propulsion has been particularly topical in recent years, with the rise of miniature artificial swimmers as a new tool for flow control, low Reynolds number mixing, micromanipulation or even drug delivery.
It is possible to take advantage of interfacial physics to propel these micro-robots, as demonstrated by recent experiments using the proximity of an interface, or the interface itself, to generate propulsion at low Reynolds number.
This paper discusses how a nearby interface can provide the symmetry breaking necessary for propulsion.
An overview of recent experiments illustrates how forces at the interface can be used to generate locomotion.
This paper then presents original results concerning two systems.
The first is composed of floating ferromagnetic spheres that assemble through capillarity into swimming structures.
The second system, also powered by a magnetic field, is a centimeter-sized piece that swims similarly to water striders.
} 
\maketitle


\section{Introduction}

An interface, the frontier between two media, is often a region of interest for scientists.
In parti\-cular, surface physics becomes more relevant the smaller the scale, as volume forces get weaker and weaker compared to surface forces.
In a fluid, another effect of a smaller scale is that the relative importance of inertia over viscosity decreases~\cite{batchelor2000}.
With this in mind, one could wonder what role interfaces can play in viscosity-dominated microscopic flows, and in particular how surface forces affect the locomotion of microorganisms and the swimming microrobots that mimic them.
This paper will first present the generalities of microscopic locomotion in a fluid, and what becomes of these principles when the presence of an interface is taken into account.
Then, several experiments from the literature will be discussed that use surface effects for locomotion.
Lastly, we will describe in more details two experiments of surface swimmers powered by external magnetic fields.
This last section will contain original experimental and theoretical results.

\section{Scallops, interfaces and symmetry}
\subsection{Swimming in the bulk}

We generally have a good enough intuition about what it means to swim in a fluid at our scale.
Water is pushed away in a given direction, and motion ensues in the opposite direction by conservation of momentum~\cite{taylor1951}.
Sustain the motion by repeating this periodically and we obtain a working swimming strategy.
However, as is often the case in nature, the physics of swimming is highly influenced by the relevant length and time scales~\cite{taylor1951,purcell1977,lauga2009}.
Consider the general case of a body that is able to actively deform, moving in an infinite fluid volume.
The conservation of momentum at each point in the fluid is described by the Navier-Stokes equation~\cite{navier1823,stokes1845} which, for an incompressible fluid, submitted to gravity, of density $\rho$ and kinematic viscosity $\nu$, reads
\begin{equation}
\frac{\partial \textbf{u}}{\partial t} + \textbf{u} \cdot \nabla \textbf{u} = -\frac{1}{\rho}\nabla p + \nu\nabla^2 \textbf{u} + \textbf{g}.
\label{NS}
\end{equation}
This equation must be completed with the continuity equation $\nabla \cdot \textbf{u} = 0$ and with the appropriate boundary conditions, taking into account the position at all times of the surface of the deformable body.
For example, no-slip boundary conditions stipulate that $\textbf{u}_{\mathrm{fluid}} = \textbf{u}_{\mathrm{body}}$ at each point on the surface of the body.

It is often appropriate to compare the magnitude of the various terms in equation~(\ref{NS}), in order to identify the relevant effects and provide some simplification.
Let $L$ and $U$ be a typical length and a typical speed of the flow, respectively.
For example, this could be the body length and speed of the swimmer, although it is sometimes more useful to look at the dimensions associated with what is producing the flow, like a beating fin or a rotating flagellum. 
The left member of equation~(\ref{NS}) represents inertial forces and contains the unsteady term $\partial \textbf{u} / \partial t$ and the advection term $\textbf{u} \cdot \nabla \textbf{u}$.
Both terms have the units of $U^2 /L$.
The viscous forces per unit mass $\nu \nabla^2 \textbf{u}$ scale like $\nu U / L^2$.
Therefore, the ratio of inertial and viscous forces in the flow is given by the Reynolds number Re~$=UL/\nu$~\cite{stokes1851,reynolds1884}.
This means that, for a given liquid, viscous forces tend to dominate over inertia at small scales.
Water has a kinematic viscosity of about $\nu=10^{-6}$~m$^2/$s, so that for a swimmer moving at one body length per second, \emph{i.e.} $L/U = 1$~s, we would have Re $<$ 1 for $L<1$~cm.

\begin{figure}
\includegraphics[width=\linewidth]{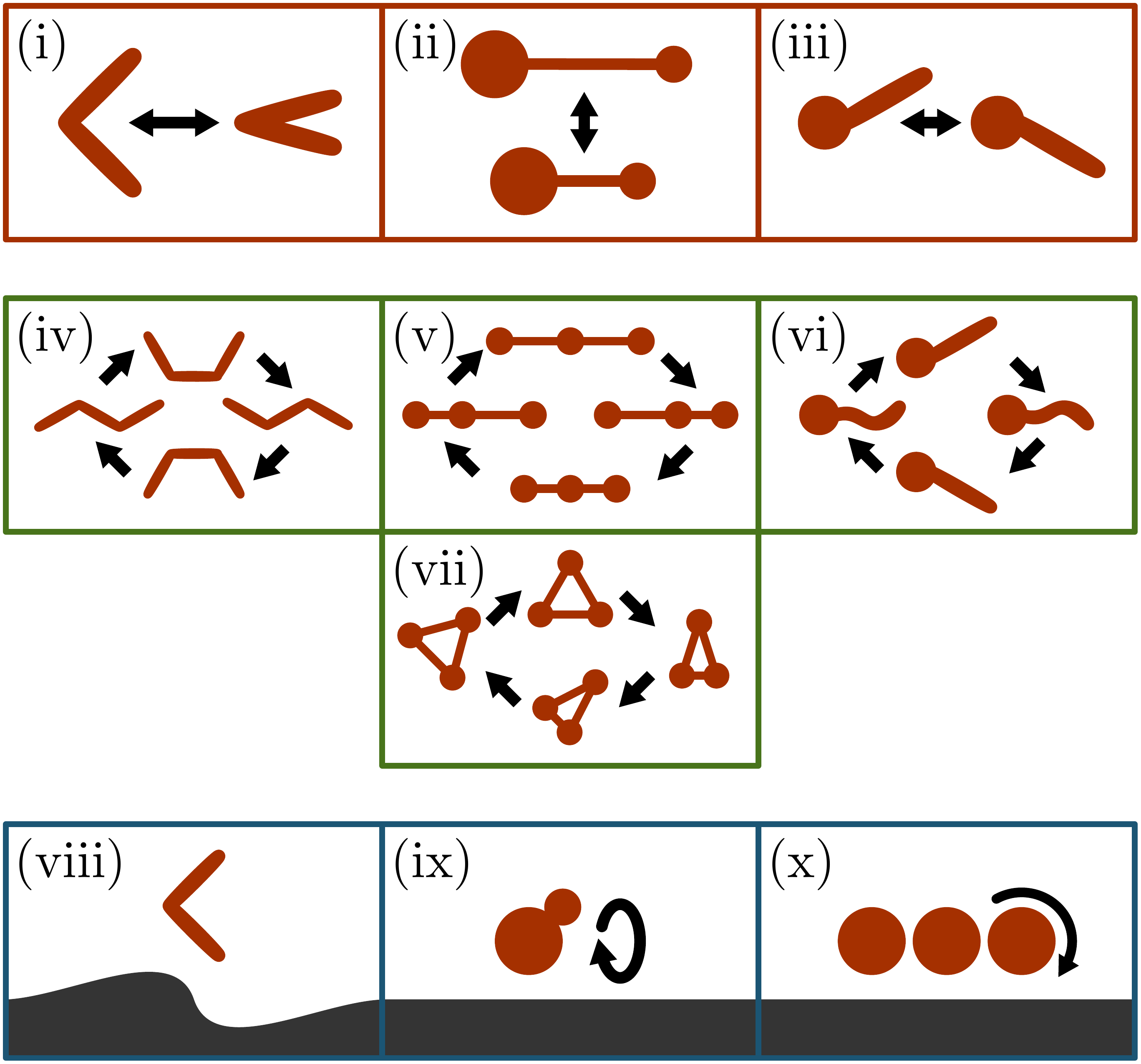}
\caption{On top, reciprocal swimming strategies that do not work in the bulk at low Reynolds number.
This includes (i)~a scallop-like swimmer~\cite{purcell1977}, (ii)~two oscillating spheres of different sizes~\cite{klotsa2015}, and (iii)~a body with a rigid beating tail~\cite{alben2005}.
In the middle, non-time-reversible deformation sequences that can produce a net motion in the Stokes regime.
The three-link swimmer~(iv) has two hinges that move out of phase~\cite{purcell1977}.
The two arms of the three-linked-spheres swimmer~(v) also oscillate out of phase~\cite{najafi2004}.
A deformable body such as a flexible magnetic tail~(vi) can also produce a non-reciprocal motion~\cite{dreyfus2005}.
Another way to propel three spheres is to use a triangular configuration~(vii), where the oscillation of one pair is accompanied by an out-of-phase rotation~\cite{grosjean2015}.
Below, swimming srategies that work in the Stokes regime only with a nearby interface.
A reciprocal swimmer such as the scallop-like one~(viii) can move in all directions when close to a deformable interface~\cite{trouilloud2008}.
Two stacked spheres in a precession movement~(ix) can move with an interface nearby~\cite{tierno2008b}.
Rotating spheres can also move close to an interface~(x), where they self-assemble into a colloidal conveyor belt under a combination of rotating and oscillating fields~\cite{martinez2015}.}
\label{scallop}
\end{figure}

For Reynolds numbers close to zero, the left member of equation~(\ref{NS}) can be neglected.
This leads to the Stokes equation
\begin{equation}
-\frac{1}{\rho}\nabla p + \nu\nabla^2 \textbf{u} + \textbf{g} = 0
\label{Stokes}
\end{equation}
which is linear and independent of time.
This has some serious consequences on the swimming mechanisms of microorganisms~\cite{purcell1977,lauga2009}.
The fact that time does not intervene in the Stokes equation means that flows are typically reversible and rate independent.
Let a body composed of two segments linked by a hinge, as shown in Figure~\ref{scallop}(i).
The opening angle between the segments is the only degree of freedom.
At high Reynolds number, this simple structure can swim by rapidly closing, expelling water, and then reopening slowly.
This is a swimming strategy similar to what some scallops do, with valves in place of the segments, except that the water is expelled through small openings on either side of the hinge.
However, the rate of closing and opening does not influence the flow in the Stokes regime, meaning that only the succession of geometric configurations adopted by the swimmer matters.
With only one degree of freedom, our model scallop can only go back and forth between the open and close configurations.
Even if water is pushed during the closing phase, it will always produce the inverse flow by reopening, so that the center of mass is not displaced over one period.
This leads to what is colloquially known as the ``scallop theorem'', stating that, at Re~$=0$, if the succession of configurations adopted by the swimmer is unchanged by a time-reversal transformation, then it cannot produce a net motion~\cite{purcell1977}.
Several other swimming strategies that have been proven to work at higher Reynolds number will fail for the same reasons.
For instance, two spheres of different sizes linked by an oscillating spring, as shown in Figure~\ref{scallop}(ii), would not be able to swim in the Stokes regime, as there is only one degree of freedom for deformation.
Such a swimmer has been shown to produce virtually no net motion under a critical value for the Reynolds number, at around Re~$\approx 20$~\cite{klotsa2015}.
Similarly, a beating rigid tail, shown in Figure~\ref{scallop}(iii), can produce no net motion if the Reynolds number is smaller than unity~\cite{alben2005}.
Depending on the particular case, the onset of a net motion for a reciprocal swimmer, following an increase in the Reynolds number, can occur either continuously or discontinuously~\cite{lauga2007}.

Another way to discuss the implications of equation~(\ref{Stokes}) is to consider that a swimmer, in the Stokes regime, is incapable of exerting a net force, or conversely, a net torque, on the surrounding fluid~\cite{lauga2009}.
Indeed, considering that inertia is negligible when Re~$\approx 0$ is equivalent to stating that the swimmer experiences a resultant force from the fluid equal to zero at all times, granted that there is no external force pushing the swimmer.
If the flow field around the swimmer is expressed as a multipole expansion, the term decaying like $1/r$, which corresponds to a point force and is called a stokeslet, is therefore zero.
The leading term in the far-field flow is thus a symmetric force dipole at best, \emph{i.e.} the flow generated by two opposite point forces, which decays like $1/r^2$.
This is called a stresslet and is a useful tool for describing in general terms the motion of a microswimmer and its interactions with its environment~\cite{lauga2009,trouilloud2008,pushkin2013}.
In general, a swimmer generates a flow that is a combination of stresslets and higher order terms, such as the source dipole, whose velocity field decays like $1/r^3$.

For decades now, researchers have come up with swimming strategies that satisfy the conditions imposed by the scallop theorem~\cite{purcell1977,lauga2009,najafi2004,dreyfus2005,tierno2008,grosjean2015}.
Such strategies already exist in nature and are used by motile bacteria and sperm cells.
This includes the rotation of one or several helical flagella~\cite{purcell1977}, the sequential motion of a series of cilia~\cite{lauga2009} or the transfer of mass by deformation of the whole cell membrane~\cite{farutin2013}.
However, it is possible to devise strategies that are conceptually simpler, more adapted to analytical calculations or numerical simulations, and/or easier to implement experimentally with existing technologies.
One early example, proposed by Purcell in 1976~\cite{purcell1977}, is the addition of one degree of freedom in the scallop-like system from Figure~\ref{scallop}(i), which is now composed of two hinges and three segments, the so-called three-link swimmer, as shown in Figure~\ref{scallop}(iv).
This allows to move the external arms one after the other, leading to a sequence of configurations that is not time-reversible.
As shown, the sequence leads to a net motion to the right.
This is easier to understand when picturing this swimmer as standing on a sinusoidal wave travelling to the left, where every configuration change moves the wave by a quarter wavelength.
Experimental implementations of this swimmer have been made, though they require macroscopic elements such as motors.

An arguably simpler, one-dimensional deformation sequence has been proposed by Najafi and Golestanian in 2004~\cite{najafi2004}.
Like the swimmer from Figure~\ref{scallop}(ii), it consists of spheres linked by arms whose length can vary.
In order to beat the scallop theorem, a minimum of three spheres and two independent arms is required.
The sequence as depicted in Figure~\ref{scallop}(v) leads to a net motion to the right.
A lot of theoretical work has been based on this swimmer, sometimes using oscillating springs instead of arms~\cite{golestanian2008,zargar2009,pickl2012,pande2015,pande2017}.
This can be attributed in part to its one-dimensional nature which greatly facilitates analytical calculations.
Notably, it can be shown that the speed of this swimmer over one period is proportional to the area of the cycle drawn by the swimmer in the plane defined by the two arms' lengths~\cite{golestanian2008}.
If the arms oscillate harmonically at a frequency $\omega$, this can be expressed as the product of the amplitudes of oscillations of each spring with the sine of their phase difference $\phi$, namely
\begin{equation}
V = K A_1 A_2 \, \omega \sin \left( \phi \right) = K W,
\label{golestanian}
\end{equation}
where $K$ is a geometrical prefactor that can be determined analytically, and we defined swimming efficiency $W$.

Despite its simplicity, the three-linked-spheres swimmer is far from being the first to have been implemented experimentally.
This honour goes to the work by Dreyfus \emph{et al.} in 2005, which is often regarded as the first artificial microswimmer~\cite{dreyfus2005}.
A magnetic tail, composed of superparamagnetic colloids linked by DNA strands, is attached to a red blood cell.
In an oscillating magnetic field, the tail deforms and aligns itself periodically with the field, following a non-time-reversible sequence illustrated in Figure~\ref{scallop}(vi).

Another possible deformation sequence uses three spherical particles forming a regular triangle, as shown in Figure~\ref{scallop}(vii)~\cite{lumay2013,grosjean2015}.
It only requires one pair of spheres to oscillate, while the other two arms can remain rigid.
In this case, the rotation of the ensemble is the key ingredient to generate a non-reciprocal cycle.
Indeed, the center of rotation is determined by the hydrodynamic interaction between the spheres.
During each contraction of the pair, it moves away from the center of mass, which is then displaced by the rotation.
Once the swimmer goes back to the equilateral configuration, the center of mass and the center of rotation are confounded again, so that a net displacement has been produced over one cycle.
Compared to the one-dimensional three-bead-swimmer, the triangular one can freely move in the plane with two degrees of freedom.

\subsection{Beyond the scallop theorem}

While the scallop theorem has been the basis for many studies, there are several cases where it is not applicable.
For instance, the independence in the rate of deformation of the body is only valid in a Newtonian fluid.
In a shear thinning or shear thickening fluid, for example, the change in apparent viscosity can be used to produce a net displacement with a reciprocal motion~\cite{lauga2009b,qiu2014}.

The proximity of another body can also be used to relax the condition imposed by the scallop theorem.
For instance, two out-of-phase reciprocal swimmers can essentially act as one non-reciprocal swimmer~\cite{lauga2008}.
A nearby interface, which is a common scenario in biological fluids or microfluidic devices, can also be used to beat the scallop theorem.
In their 2008 paper, Trouilloud \emph{et al.} studied the flow induced by a reciprocal swimmer near an interface, by looking at the flow in the far field as a superposition of stresslets and source dipoles~\cite{trouilloud2008}.
In this case, while the proximity of a rigid wall can induce an additional velocity component, it does not allow to overcome the scallop theorem.
However, a reciprocal swimmer can move when the interface in question is deformable, such as the interface between two fluids, or between a fluid and a deformable solid, like a membrane or a gel.
Swimming is possible towards, away and parallel to the interface, depending on the stresslets and source dipoles considered.
To generate a significant motion, the swimmer must be able to generate an important enough deformation of the interface.
The swimmer exerts a typical viscous force $\eta U L$ on the interface, where $L$ represents both the swimmer size and its distance to the interface, which must be compared to a typical restoring force, such as a capillary force $\gamma L$ in the case of an interface between two fluids.
In this case, one obtains the capillary number $\eta U/\gamma$ which must be larger than unity while keeping the Reynolds number small.
This leads to a typical length scale $\eta^2 / \rho \gamma$, the Ohnesorge length, under which this kind of propulsion is effective.
Note that for simplicity, two fluids of similar viscosity and density were considered.

While a deformable interface is required in the work by Trouilloud \emph{et al.}, it is possible to generate motion parallel to a rigid interface with reciprocal motion.
In experiments performed by Tierno \emph{et al.}, a small paramagnetic sphere is attached to a larger one~\cite{tierno2008,tierno2008b,tierno2010}.
The doublet is then submitted to a precessing magnetic field, as illustrated in Figure~\ref{scallop}(ix).
Motion is induced in a direction perpendicular to the precession axis and parallel the wall.
Here, the presence of the wall creates a difference in viscous dissipation between each part of the rotation, far and close to the interface.
Indeed, the ``return stroke'' experiences a lower viscous dissipation than the ``forwards stroke'', which is closer to the interface.
This could be compared with the previously discussed reciprocal motion in a non-Newtonian fluid, except that the modulation in viscous dissipation is due to the proximity of the interface, and not the fluid itself.
This asymmetry explains the apparition of a net motion that would not be observed in the bulk.

Similarly, a simple rotational motion in a plane perpendicular to a nearby interface can lead to propulsion, as the flows below and above the rotating body differ.
On the other hand, several magnetic colloids rotating in a plane parallel to the interface can self-assemble into a two-dimensional colloidal carpet, as the time-averaged dipole-dipole interaction between the spheres is an attraction.
By combining the two effects, it is possible to generate a colloidal carpet that moves along the interface~\cite{martinez2015}.
This is illustrated in Figure~\ref{scallop}(ix).
The speed of the carpet increases with the number of particles and saturates around $N \approx 300$.
This structure can work as a conveyor belt able to transport a cargo.

\section{Swimming at the interface}

\begin{figure}
\includegraphics[width=\linewidth]{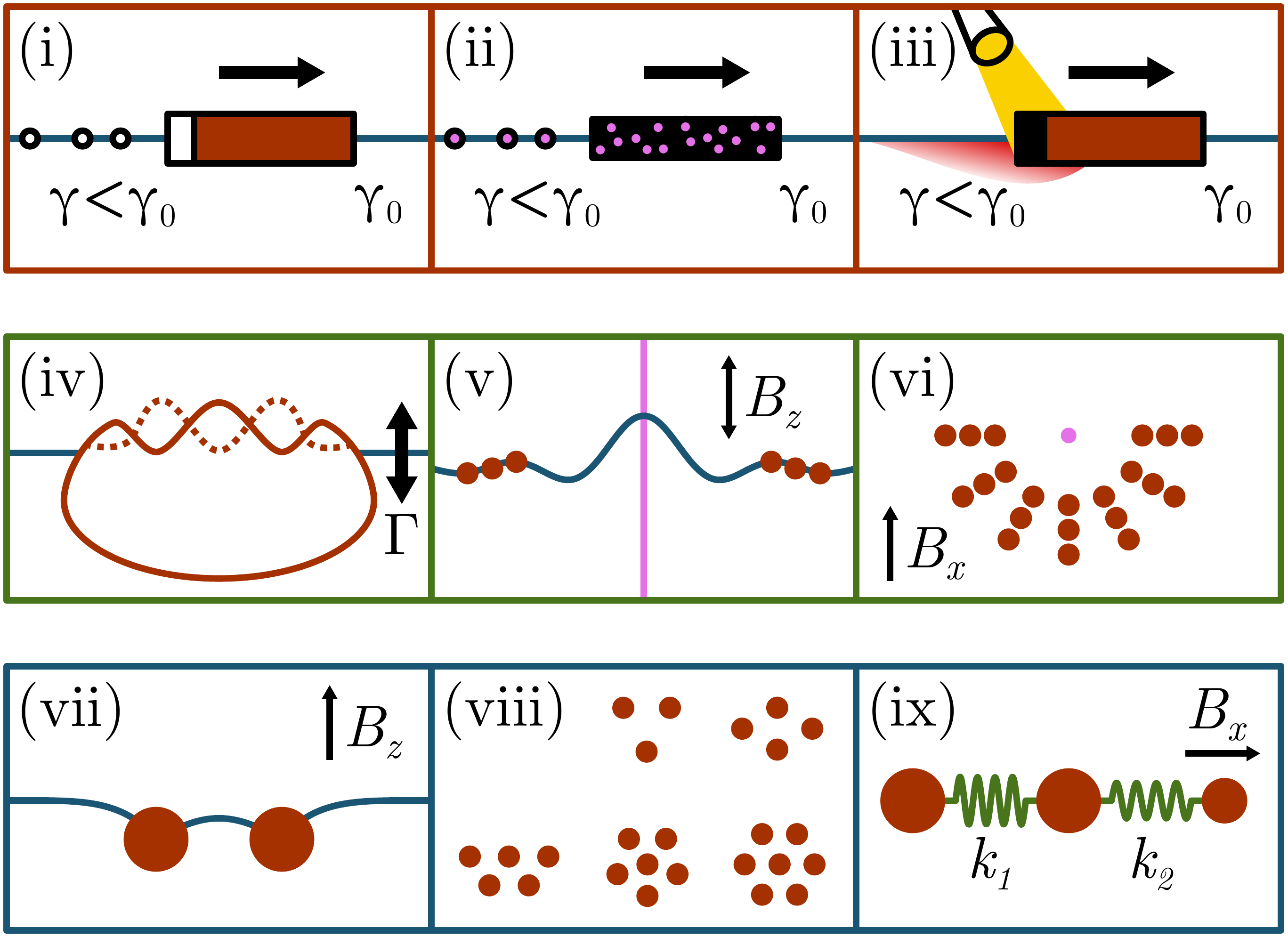}
\caption{On top, swimmers that propel using a gradient in surface tension, also known as the Marangoni effect.
In the classic camphor boat~(i), surface tension is locally lowered by a piece of camphor dissolving at the stern~\cite{nakata2013}.
Marangoni-driven propulsion can also be obtained by releasing a solvent~(ii) contained in a gel~\cite{bassik2008} or in a droplet coated with colloids~\cite{bormashenko2015}.
Note that motion can arise even with symmetric objects, as a spontaneous breaking of symmetry can be observed.
The gradient in surface tension can also come from a temperature change~(iii), for instance using a light source~\cite{okawa2009}.
In the middle, two systems that use surface waves for propulsion.
A droplet placed on a vertically vibrating bath can be deformed by a Faraday standing wave~(iv), leading to a net motion after a symmetry breaking in the position of the nodes~\cite{ebata2015}.
Magnetic colloids~(v) can also generate surface waves under a vertical oscillating field~\cite{snezhko2006}.
The particles arrange to form an aster~(vi), which can swim if the spatial symmetry is broken~\cite{snezhko2011}.
Below, floating magnetic spheres under a vertical constant field~(vii) arrange into structures~(viii) due to a competition between magnetic and capillary forces~\cite{lumay2013}.
These assemblies can move under oscillating fields~(ix), for example by mimicking the deformation sequence of the three-linked-spheres swimmer from Figure~\ref{scallop}(v)~\cite{grosjean2016}.}
\label{surface}
\end{figure}

In the previous section, we discussed how the breaking of spatial symmetry provided by a nearby interface can allow to beat the scallop theorem.
We will now show that interfacial phenomena, such as the Marangoni effect, surface waves or the so-called Cheerios effect, can also be used to generate microswimmers.

\subsection{Marangoni effect}

In the scallop-theorem paradigm, it is assumed that the self-propulsion of the swimmer is achieved through the deformations of the body.
However, it is also possible for a rigid body to achieve a force-free self-propulsion through a self-generated gradient.
This is the basic principle behind self-phoretic swimmers, which induce flows on their surface through gradients in concentration, temperature or electrostatic potential~\cite{golestanian2007}.
For example, self-diffusio\-phoresis can arise when a particle is partially covered with a catalyst for a chemical reaction that can occur in the surrounding fluid, locally creating a gradient in concentration~\cite{howse2007,popescu2016}.
The asymmetry is not necessarily required, as a symmetry breaking can also spontaneously occur with isotropic particles~\cite{michelin2013}.
The process is similar with self-thermo\-phoresis~\cite{jiang2010} or self-electro\-phoresis~\cite{pumera2010}.
Self-phoretic swimmers are in a class of their own, with many mechanisms involved, which is why it will not be discussed in more detail here.
However, the principle of a self-generated gradient is also used for motion along an interface, in the case of propulsion by Marangoni effect.
The camphor boat, which has been known for more than a century, is now used as a model system for low Reynolds locomotion~\cite{nakata2013,karasawa2014}.
A piece of camphor is attached to a floating object.
When it dissolves in the water, the camphor molecules adsorbed at the water surface locally lower surface tension, as illustrated in figure~\ref{surface}(i).
The resulting surface tension gradient propels the object forwards.
While the camphor boat is the earliest and most well-known example of propulsion by Marangoni effect, it is far from being the only one.
For example, a body releasing a solvent in the surrounding liquid can generate a gradient in surface tension, as represented in Figure~\ref{surface}(ii).
Examples of bodies placed on a water bath include a gel disk soaked in oxolane (tetrahydrofuran)~\cite{mitsumata2001} or ethanol~\cite{bassik2008}, a droplet of aqueous ethanol coated with colloidal particles, called a liquid marble~\cite{bormashenko2015}, and a soap disk at an oil-water interface~\cite{nakata2005}.
This effect has also been observed with pure water droplets placed on an oil-surfactant bath~\cite{izri2014}.
Note that these objects can be isotropic, as any small anisotropy due to the initial conditions can increase when the object starts to move.
A variation on this principle is to generate a surface tension gradient by locally heating the fluid, for example by illuminating an object with intense light.
This is illustrated in Figure~\ref{surface}(iii), where a light-absorbing element is placed at the back of an object heated with focused light~\cite{okawa2009}.
Using a laser as heat source makes it possible to move isotropic objects such as steel spheres, as it allows light to hit the surface at a precise point~\cite{mallea2017}.
Finally, stationary heated structures on a chip suspended above the surface can generate many types of behaviours by using point, line, annular or triangular heat sources~\cite{basu2007}.

\subsection{Surface waves}

Another type of interfacial phenomenon that can be used for propulsion is surface waves.
For instance, one can use Faraday waves, \emph{i.e.} standing waves that appear on a vibrating bath, for locomotion.
A famous example is the case of walking droplets, where an oil droplet bouncing on a vibrating bath, just below the onset of the Faraday instability, generates waves that help propel it forwards~\cite{couder2005}.
This does not technically qualify as swimming, as the droplet is never immersed in the bath.
It is possible, however, to produce a swimmer by using a similar system~\cite{ebata2015}.
In this case, a water droplet is placed on a vibrating bath of silicon oil.
The droplet is mostly immersed, with a small cap peaking above the surface of the bath.
Under a strong enough acceleration of vibration $\Gamma$, a Faraday standing wave can appear on the surface of the water droplet, as depicted in Figure~\ref{surface}(iv).
This generates a flow in the surrounding oil that can lead to motion.
Depending on the forcing parameters, several types of motions are observed, including spinning, rotation on an orbit, zig-zag and translational motion.
This is linked to the number and positions of the nodes of the standing wave on the droplet.
Indeed, if the nodes are in a straight line, the droplet is either stationary or spinning.
Conversely, for some values of the forcing parameters, a symmetry breaking in the position of the nodes can spontaneously appear, leading to a net motion.
The Reynolds number for this system is typically around 0.1.
With a less viscous bath, and thus a higher Reynolds numbers of around 10, another type of motion can be observed.
In this case, the wave on the droplet is a travelling wave, leading to locomotion in the opposite direction.
This resembles the squirming model for microswimmers, where a sphere deforms its surface to generate a flow, similarly to what is observed with some bacteria such as ciliates~\cite{pak2014}.

In a second example, ferromagnetic particles floating on water can produce waves when submitted to a vertical oscillating field $B_z$.
Indeed, while the interaction between vertical dipoles on the surface is repulsive, nearby particles can form chains with a resulting moment in the plane of the interface.
These chains can deform the interface as they try to align themselves with the vertical field, which leads to the formation of self-organized structures, such as ``snakes''~\cite{snezhko2006} and asters~\cite{snezhko2011}.
A side view of an aster is depicted in Figure~\ref{surface}(v).
The colloidal chains stand on the slope of the standing wave they produce.
The addition of a constant horizontal field $B_x$ can break the circular symmetry of the aster, as shown in Figure~\ref{surface}(vi).
The asymmetric aster generates a net fluid flow, leading to locomotion.
Note that, while the particles have a typical size of 90~$\mu$m, inertia is not negligible in the flow, as the Reynolds number is of the order of 10.

\subsection{Magnetocapillary swimmers}

\begin{figure}
\includegraphics[width=\linewidth]{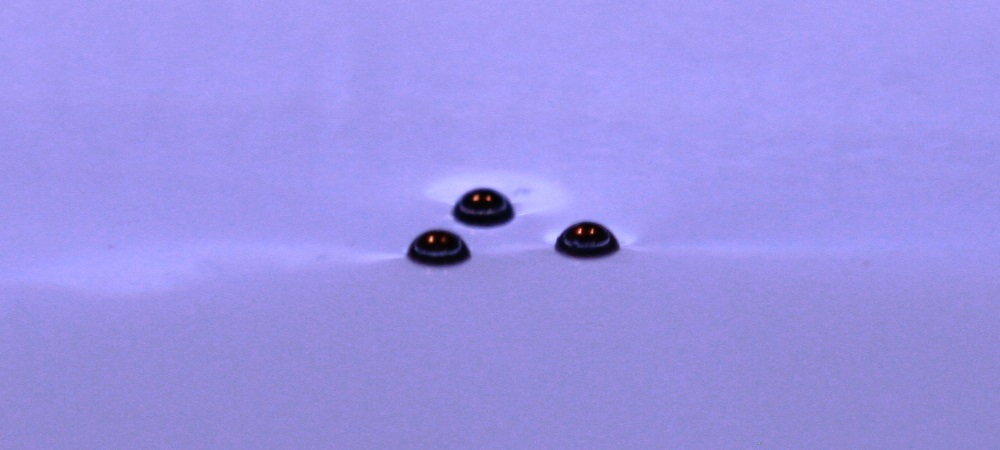}
\includegraphics[width=\linewidth]{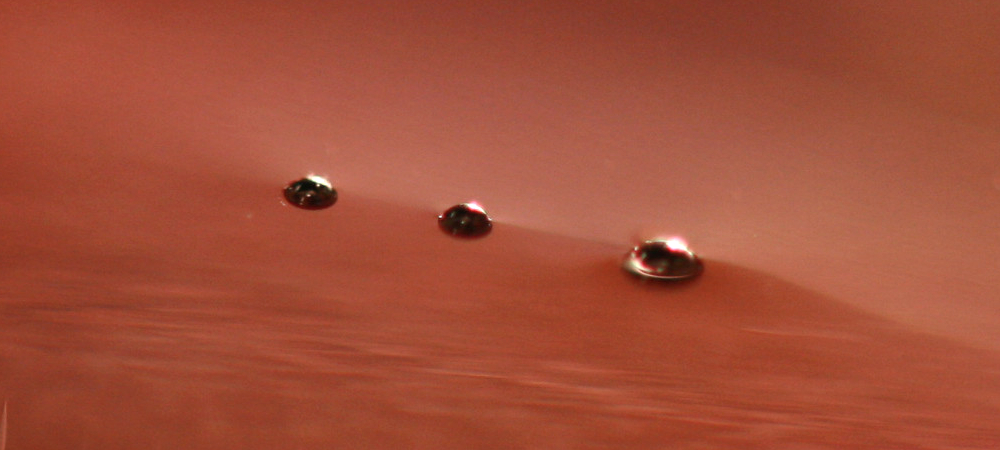}
\caption{On top, a triangular magnetocapillary swimmer, composed of three 500~$\mu$m spheres floating on water submitted to a vertical field $B_z\approx 3$~mT.
On bottom, a collinear swimmer composed of two 400~$\mu$m and one 500~$\mu$m sphere.
In this case we have a vertical $B_z\approx 5$~mT and horizontal $B_x\approx 3$~mT.}
\label{magnetocapillary}
\end{figure}

The last type of surface microswimmer that we will discuss is the magnetocapillary swimmer~\cite{lumay2013,hubert2013,grosjean2015,chinomona2015,lagubeau2016,grosjean2016,grosjean2017}.
Metallic spheres of 500~$\mu$m in diameter are placed on a water surface under a constant vertical magnetic field $B_z$.
The spheres experience an attractive force due to capillarity.
Indeed, each particle is surrounded by a meniscus, as the weight of the particle deforms the surface.
This leads to the apparition of a lateral capillary force, which is an attraction in the case of similar particles~\cite{kralchevsky1994}.
This effect is colloquially known as the Cheerios effect, as it can be observed simply with breakfast cereals in a bowl of liquid~\cite{vella2005}.
The vertical field $B_z$ is used to counter this attraction.
While they are made of a ferromagnetic material, the spheres have an almost linear magnetization due to finite size effects~\cite{lagubeau2016}.
This means that they behave essentially like paramagnets, except for a larger (effective) susceptibility $\chi_{\mathrm{eff}} \approx 3$.
However, the beads can reorient in an external field, an effect which can be attributed to a small residual magnetism of the order of 100~A/m.
Under a vertical field, the magnetic interaction between the spheres is a repulsion.
The combination of these two forces can lead to a finite equilibrium distance, as illustrated in Figure~\ref{surface}(vii).
For more than two particles, organized structures emerge, typically following a triangular lattice~\cite{lumay2013}.
The basic structures observed, up to $N=7$, are shown in Figure~\ref{surface}(viii).
A photograph of a triangular magnetocapillary swimmer is shown in Figure~\ref{magnetocapillary}.

\begin{figure}
\includegraphics[width=\linewidth]{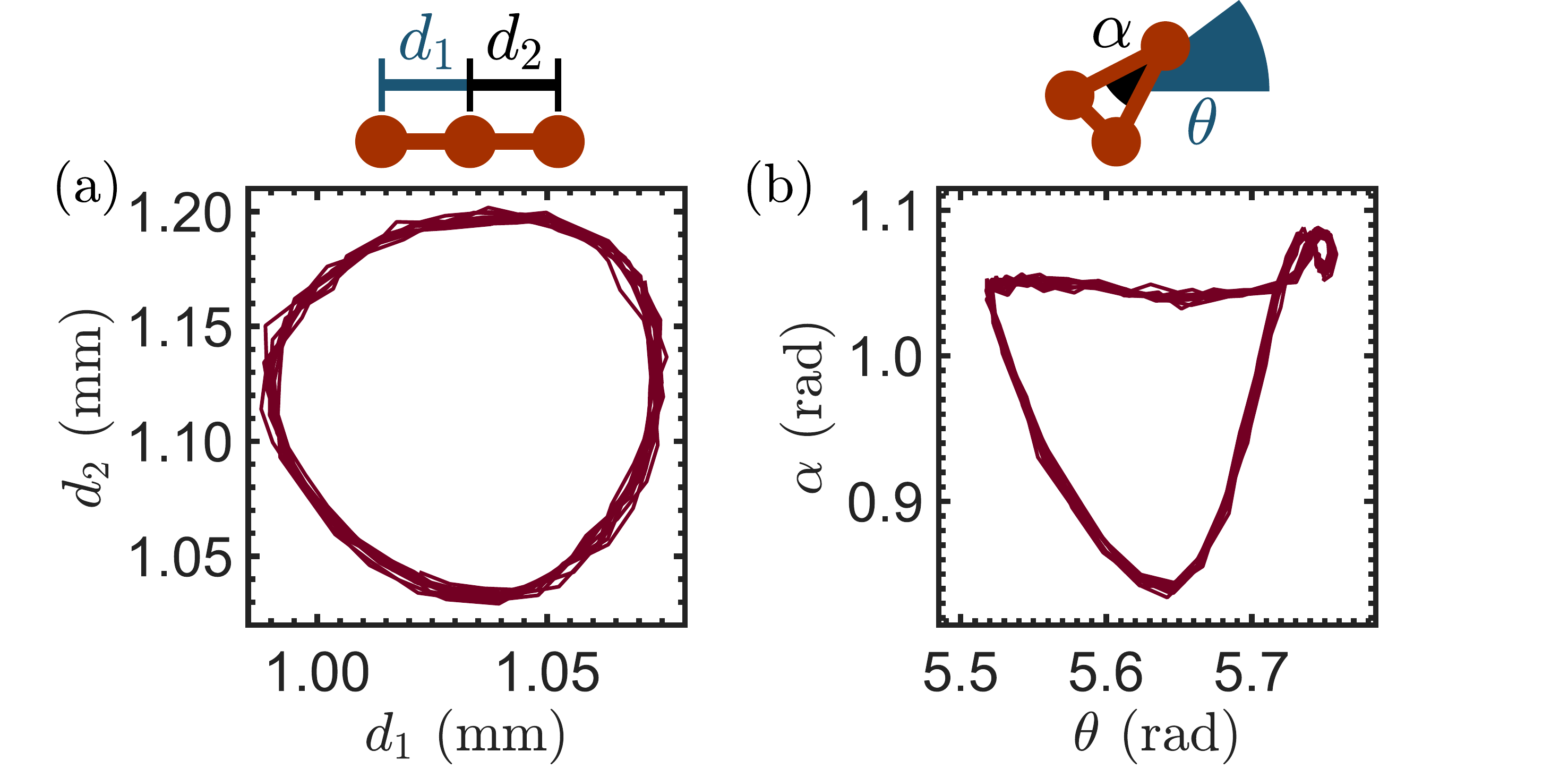}
\caption{(a) Interdistances $d_1$ and $d_2$ describe a non-reciprocal cycle in a collinear magnetocapillary swimmer.
The experiment was run for ten oscillations at 3~Hz.
(b) An internal angle $\alpha$ and the orientation of the swimmer $\theta$ also describe a non-reciprocal cycle in a triangular swimmer.
Orientation $\theta$ is defined as the average of the orientations of the three particles in the referential of the center of mass.
Ten oscillations at 0.5~Hz are shown.
}
\label{cycles}
\end{figure}

\subsubsection{Collinear swimmer}

In order to beat the scallop theorem, a minimum of three rigid spheres is required~\cite{lumay2013}, which is why the triangular swimmer was the first one to be studied in depth~\cite{grosjean2015}.
However, this triangular swimmer is neither the only, nor the most simple three-particle swimmer obtainable with a magnetocapillary system.
In fact, under a large enough constant horizontal field, a collinear configuration becomes stable~\cite{chinomona2015}.
An exemple of such a configuration is shown in Figure~\ref{magnetocapillary}.
This means that it is possible to mimic the deformation cycle of the three-link swimmer depicted in Figure~\ref{scallop}(v)~\cite{grosjean2016}.
To generate the deformation, a horizontal field of the form
\begin{equation}
B_x = B_{x,0} + \delta B \sin \left(2\pi f t\right)
\end{equation}
is used, where $\delta B \ll B_{x,0}$ in order to maintain the swimmer in the collinear state.
Identical particles will oscillate around their equilibrium position in a time-reversible way, with the magnetocapillary interaction essentially acting as a non-linear spring that brings the particles back to their equilibrium position.
In order to break time-reversal symmetry, one must introduce an asymmetry in the system.
This is simply done by changing the size of one of the external particles, as shown in Figure~\ref{surface}(ix) as well as Figure~\ref{magnetocapillary}.
In this spring analogy, this is equivalent to changing the spring constant of one of the two springs, which can introduce a phase difference between the oscillations.
The swimmer therefore follows a deformation sequence similar to the one proposed in~\cite{najafi2004} and depicted in Figure~\ref{scallop}(v).
Note that, while inertia in the flow can be neglected, the oscillations of the springs must not be overdamped in order to observe the phase shift, which means that the inertia of the particles themselves must be considered.
A typical experimental deformation cycle is shown in Figure~\ref{cycles}(a).
This approximately circular trajectory in the $(d_1,d_2)$ plane means that the two oscillations are in quadrature for $f \approx 3$~Hz.
Note that the maximum speed is not necessarily reached at the phase quadrature, as the amplitudes of both oscillations are also a function of $f$.
The amplitudes reach their maximum at the radial magnetocapillary resonance frequency described in~\cite{lagubeau2016}, which happens typically around 2 or 3~Hz.

\begin{figure}
\includegraphics[width=\linewidth]{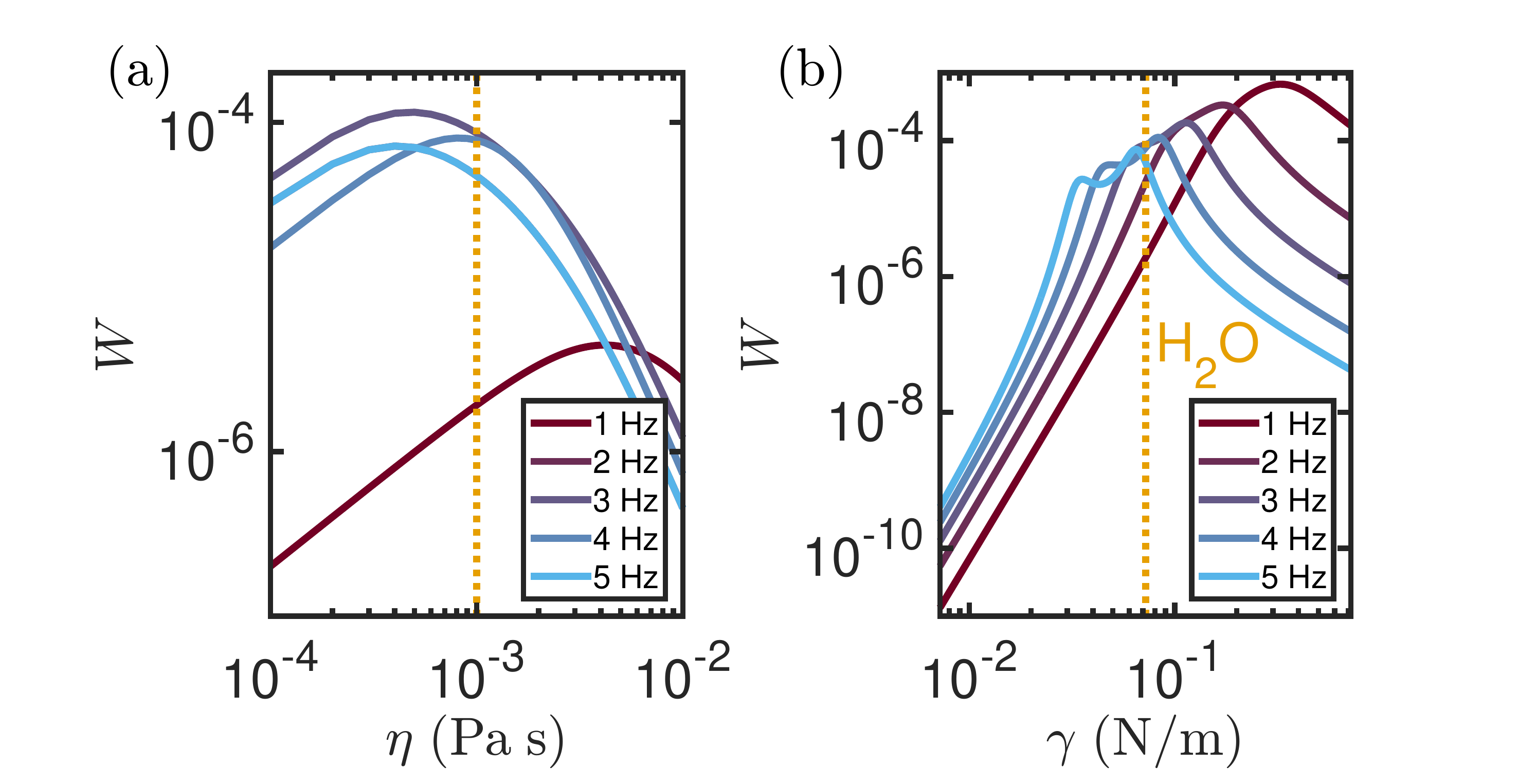}
\caption{(a) Swimming efficiency $W$ of the collinear magnetocapillary swimmer as a function of viscosity, for various values of the excitation frequency $f$.
(b) Swimming efficiency $W$ of the collinear magnetocapillary swimmer as a function of surface tension, for various values of the excitation frequency $f$.
The contact angle between the spheres and water was kept constant.
}
\label{speed}
\end{figure}

An analytical expression for the swimming speed can be expressed by combining the equations of motion of the particles with equation~(\ref{golestanian}).
Within this framework, and with the possibility of analytical developments, the collinear swimmer could be used as a model system to verify some general principles of microswimmers~\cite{pande2017,grosjean2016}.
For instance, it has been shown that the influence of the viscosity of the surrounding fluid on a swimmer is not always trivial and, in particular, that an increase in viscosity can counter-intuitively lead to an increase in speed~\cite{pande2017}.
This is also expected in the case of the magnetocapillary swimmer, as shown in Figure~\ref{speed}(a).
There seems to exist an optimal viscous damping in terms of swimming speed, which is a function of the excitation frequency $f$.
The magnetocapillary collinear swimmer could therefore offer a way to experimentally validate the results from~\cite{pande2017}.
Similarly, there is an optimal surface tension $\gamma$ for a given excitation frequency, as shown in Figure~\ref{speed}(b), where $\gamma$ has been varied while keeping the contact angle constant.
If we consider that the particles are linked by a magnetocapillary spring, the role of surface tension is essentially to act as an extension spring force while the dipole-dipole repulsion induced by $B_z$ acts as a compression spring force.

\subsubsection{Triangular swimmer}

To generate non-reciprocal motion in a triangular swimmer, it is submitted to a horizontal oscillating field
\begin{equation}
B_x = B \sin \left(2\pi f t\right)
\end{equation}
with $B<B_z/\sqrt{2}$ to avoid contact between the particles~\cite{grosjean2017}.
A constant horizontal field $B_{x,0}$ can also be added, which tends to force the swimmer into a particular swimming mode by further breaking spatial symmetry in the system~\cite{grosjean2015}.
In general, the frequency of the oscillating field is below 1~Hz, which leads to a relatively low Reynolds number for the size of the particles, usually between $10^{-3}$ and $10^{-1}$.
Higher excitation frequencies tend to lead to more complex behaviours~\cite{hubert2013}.

\begin{figure}
\includegraphics[width=\linewidth]{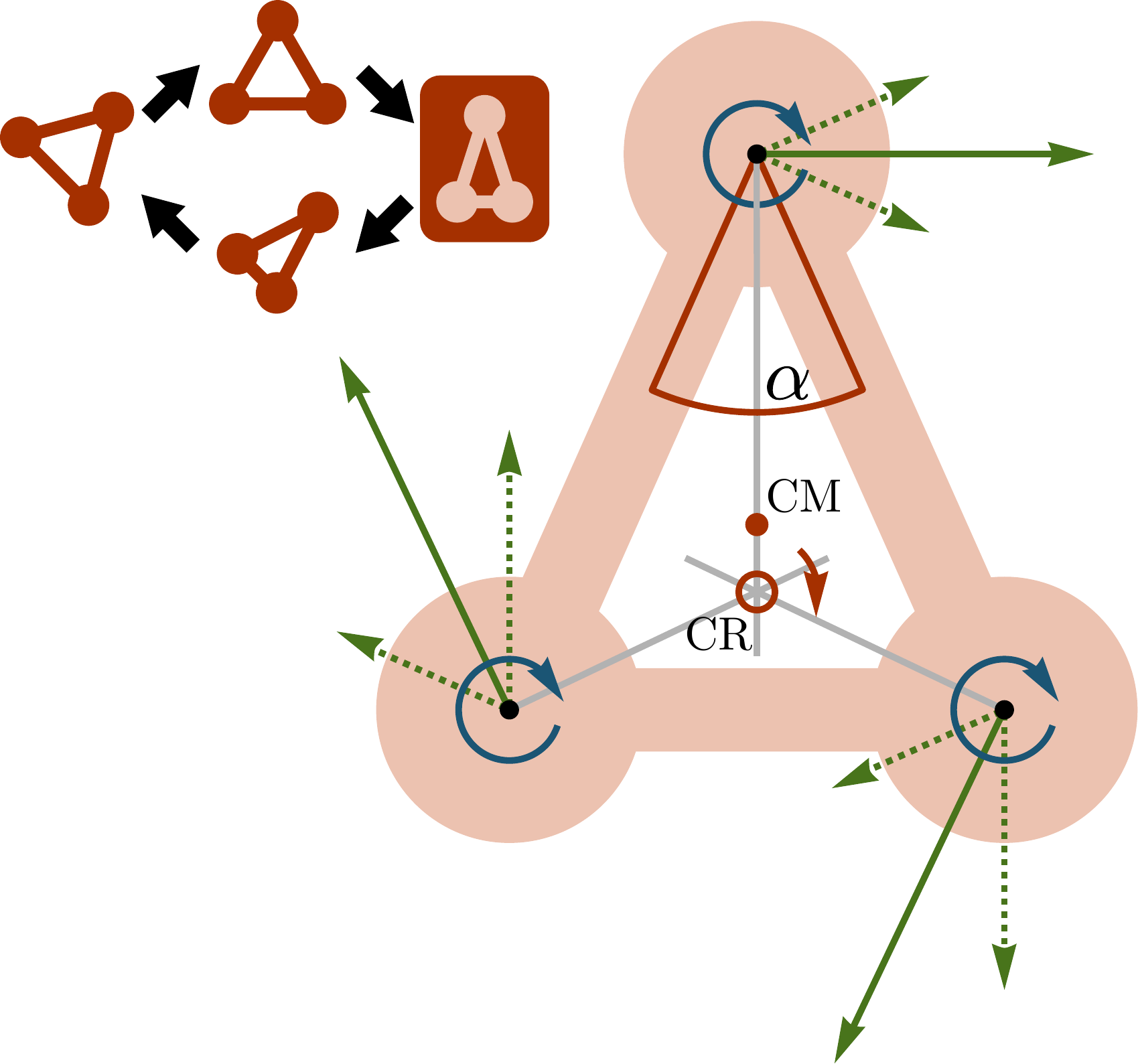}
\caption{
The magnetocapillary triangle swims thanks to a combination of deformation and rotation.
The rotation-translation hydrodynamic coupling (dotted lines), caused by the individual rotation of the spheres, defines a rotation center CR (red circle) that, in an isosceles of apex angle $\alpha$, is distinct from the center of mass CM (red dot).
Therefore, the rotation leads to a displacement of CM that is preserved if the return rotation happens when the triangle is equilateral.
CR moves closer or further away from the apex depending if $\alpha>\pi/3$ (\emph{platy}) or $\alpha<\pi/3$ (\emph{lepto}), respectively.
In the experiment, the swimmer tends to deform into a flat isosceles during the return rotation~\cite{grosjean2015}, which further increases the effect.
}
\label{lepto}
\end{figure}

One typical deformation sequence observed is similar to the one depicted in Figure~\ref{scallop}(vii).
The regular triangle deforms into an isosceles as the magnitude of the field increases.
A rotation is then observed, which is kick started by the presence of a small residual magnetism in the spheres, leading to a torque on the assembly.
The center of mass CM moves, as it is distinct from the center of rotation CR, which is determined by the hydrodynamic interactions between the spheres.
The swimmer then goes back to a less deformed state and rotates back to its initial orientation, leading overall to a net displacement of the center of mass.
Figure~\ref{lepto} illustrates this process, showing a typical cycle going from equilateral to a pointy isosceles, which then deforms back into an equilateral and rotates back to its initial configuration.
The pointy isosceles state, whose apex angle $\alpha$ is below $\pi/3$ and which is called \emph{lepto}, is shown in more detail.
The center of rotation is defined by the hydrodynamic coupling between the spheres.
Indeed, assuming each sphere rotates individually due to the magnetic field, the induced flow field leads to a force on the neighbouring particles.
The resulting forces define a rotation center that is aligned with CM and CR.
In the \emph{platy} case, CM is closer to the apex than CR.
In the case of a flat isosceles, whose apex angle $\alpha$ is above $\pi/3$ and which is called \emph{platy}, CM would be further away from the apex than CR.
The \emph{lepto} and \emph{platy} configurations were observed in both experiments and quasi-static simulations, though with a different particle at the apex~\cite{grosjean2015}.
Their effect in the cycle is complementary, as they rotate in opposite directions.
For the sake of simplicity though, the cycle from Figure~\ref{lepto} only goes back and forth between a regular and a \emph{lepto} configuration, as the \emph{platy} state is less pronounced and not essential to explain the non-reciprocal cycle.
An experimental deformation cycle is shown in Figure~\ref{cycles}(b).
The two degrees of freedom shown are the internal angle $\alpha$, corresponding to the apex of the \emph{lepto} triangle, and the orientation of the triangle $\theta$, defined as the average orientation of the spheres in the referential of the center of mass.

Because the particles are further away from each other on average and due to the restrictions on the amplitude $\delta B$, the collinear swimmer is usually about ten times slower than the triangular one, whose typical speed is around one particle radius per period of the oscillating field.
This is why the triangular swimmer, despite being more complex, is more appropriate in the study of potential applications of microswimmers.
Indeed, it is possible to control rather precisely its trajectory in the plane of the interface~\cite{grosjean2015}.
The possibility of capturing, transporting and releasing a floating cargo has been demonstrated experimentally, as well as the mixing of fluids at low Reynolds number~\cite{grosjean2017}.

\subsection{Surface effects at larger scales}

\begin{figure}
\includegraphics[width=\linewidth]{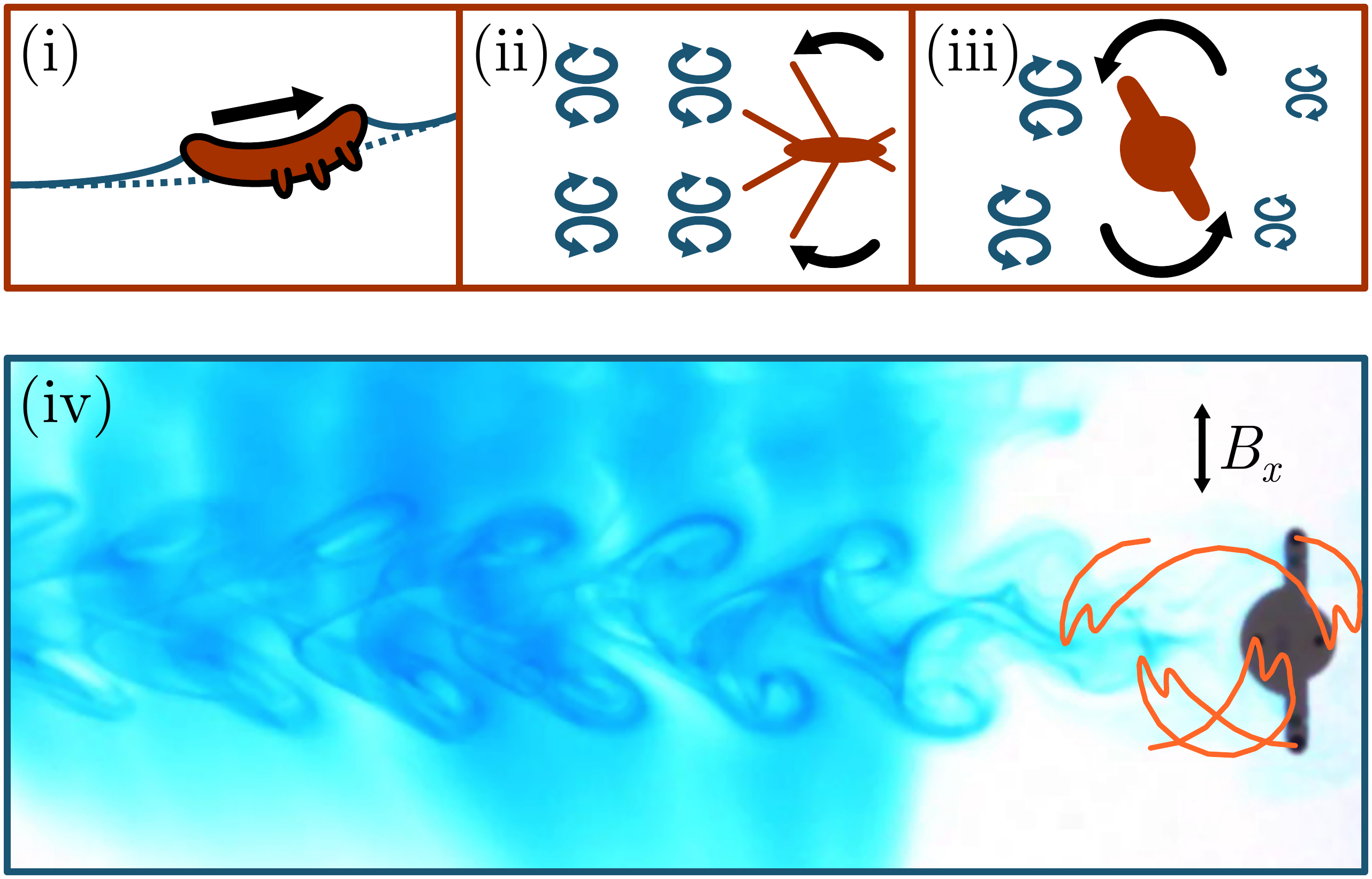}
\caption{
Exemples of centimeter-scale surface locomotion.
The larva of the waterlily leaf beetle (i) can move upwards a meniscus by deforming to generate a Cheerios effect.
Water striders (ii) float and propel vortices thanks to their hydrophobic legs~\cite{hu2003}.
A simple surface swimmer (iii) composed of two arms and an asymmetric body can swim by expelling vortices.
There is a net motion to the right thanks to the asymmetry in momentum transfer to the fluid.
Experimentally (iv), the oscillation of the piece is achieved thanks to embedded magnets in the arms and an external oscillating field.
}
\label{meso}
\end{figure}

While the systems described in this paper mainly belong to the realm of low-Reynolds-number flows, it should be noted that interfacial effects are typically relevant up to the centimeter scale.
In fact, the capillary length $l_c = \sqrt{\gamma/\rho g}$, below which surface tension dominates over gravity, is around 2.7~mm in water.
This explains why some insects and other invertebrates rely on surface forces for propulsion.
For instance, some small animals can use the Cheerios effect to ascend a meniscus~\cite{hu2005}.
Water treaders, a semiaquatic insect, achieve this by pulling the interface upwards with their legs.
Alternatively, some terrestrial insects such as beetle larvae can bend their whole body to generate the same effect, allowing them to reach land after an unintended fall onto water.
Figure~\ref{meso}(i) depicts a beetle larva climbing a meniscus.
Millimeter-long nematodes, also called roundworms, have been shown to not only climb a meniscus, but aggregate and remain grouped together thanks to the Cheerios effect~\cite{gart2011}.

Other invertebrates, such as water striders~\cite{hu2003} and fisher spiders~\cite{suter1997}, float on water thanks to surface tension.
They use their hydrophobic legs to transfer momentum to the liquid by generating U-shaped vortex rings attached to the interface, as represented in Figure~\ref{meso}(ii).
This type of locomotion relies on a higher Reynolds number, typically around 100 or more~\cite{hu2003}.
To achieve this, water striders possess three pairs of legs that secrete a hydrophobic wax.
They are covered with microscopic needles, called setae, which themselves are marked with a multitude of nanogrooves~\cite{feng2007}.
Only the middle pair of legs is used for propulsion.
This motion resembles rowing, as the return stroke happens outside of water.
In addition, some aquatic insects such as riffle bugs, smaller relatives of the water striders, can secrete surfactants to move by Marangoni effect~\cite{bush2007}.
This is based on the same principle that was depicted in Figure~\ref{surface}~(ii).
It generates a fast motion that is used as an escape mechanism.

It is possible to design artificial surface swimmers based on similar mechanisms.
For instance, artificial water striders have been built using an elastic thread~\cite{hu2003}, piezoelectric actuators~\cite{song2007} or small dc motors~\cite{zhang2011} to power the legs.
A larger number of supporting legs can allow such robots to support heavier loads~\cite{song2007,zhang2011}.
Using 3D-printing technology, we designed a very simple structure that captures the basics of this swimming strategy, \emph{i.e.} floating on water and transferring momentum to the fluid by producing vortices.
This swimmer is composed of a body that has the shape of a disk, with a pair of arms attached, as shown in Figure~\ref{meso}(iii).
However, to generate propulsion, the fore-aft symmetry of the piece must be broken.
Therefore, the central disk has a small radius on one side (fore) and a larger radius on the other (aft).
In the example shown in Figure~\ref{meso}(iv), the arm reach of the piece is 2~cm and the radii of the body are 0.35 (fore) and 0.50~cm (aft).
When this object oscillates, it can generate vortex half-rings on each side.
Thanks to the geometric asymmetry of the piece, the expulsion of vortices itself is asymmetric.
This leads to a net motion of about 1.8~cm per period of oscillation.
Figure~\ref{meso}(iv) shows the trail left by such a swimmer in coloured water.
One can see the rather large vortices on the side with the larger radius (aft).

In practice, a permanent magnets is embedded in each arm, oriented perpendicular to it.
The oscillation of the piece is powered by an oscillating magnetic field which creates a time-dependent torque.
The field is sinusoidal, of amplitude 2.8~mT and frequency 0.5~Hz and oscillates perpendicular to the swimming direction.
A small offset of 0.28~mT is added perpendicular to the oscillation, which prevents the piece to perform a full turn.
Figure~\ref{meso}(iv) also shows the trajectory of the arms' ends over one period of oscillation.
One can see that the piece swings back and forth between $-\pi$ and $\pi$ radians.

The Reynolds number of this system is typically of a few hundreds.
Another useful dimensionless number to describe vortex shedding is the Strouhal number, often written St~$= Af/U$ where $A$ is the stroke amplitude, $f$ its frequency and $U$ is the swimming speed~\cite{batchelor2000,taylor2003}.
In the case of undulatory propulsion of fish, the optimal Strouhal number has been theoretically predicted, and ranges between 0.15 for large animals like cetaceans up to 0.8 for small animals such as tadpoles~\cite{eloy2012}.
An oscillating piece like the one in Figure~\ref{meso}(iv) has a Strouhal number of about 0.55, suggesting that vortex shedding is the relevant swimming mechanism.
Further studies could aim to optimize the efficiency of the swimming piece by varying the geometrical parameters as well as the applied field.
This could provide a model structure to study the laws that govern biolocomotion, as well as a basic element to construct untethered swimming robots.

\section{Conclusion}

It certainly makes sense, from a theoretical standpoint, to study microswimmers in an unbounded volume of fluid.
However, in real world systems such as microfluidic devices or the human body, microswimmers are highly likely to encounter obstacles, interfaces or membranes.
Here, we discussed the necessary conditions for swimming imposed by the so-called scallop theorem, which stipulates that a deforming body must adopt a non-time-reversible series of shapes in order to produce a net motion at low Reynolds number.
This condition can be relaxed in the vicinity of an interface, which can for example add an extra degree of freedom in the system.
Not only can an interface help produce the breaking of symmetry necessary for propulsion, but interfacial phenomena can also play a role in generating motion.
This includes the Marangoni effect, where a gradient in surface tension leads to a net motion; surface waves, which can generate flows in the surrounding fluid; and the Cheerios effect, where particles self-assemble into a swimmer thanks to a lateral capillary force.
The latter swimmer was studied in more depth in this paper.
Two distinct swimming mechanisms were evidenced and discussed both experimentally and numerically.
Interfacial forces can also play a role in systems with a larger Reynolds number, as seen in some insects and other invertebrates.
This was illustrated experimentally by designing an asymmetric oscillator that swims by vortex shedding.

\section{Acknowledgements}

GG thanks FRIA (FNRS) for financial support.
This work was financially supported by FNRS PDR grant T.0129.18.

\section{Authors contributions}

All the authors were involved in the preparation of the manuscript.
All the authors have read and approved the final manuscript.

\vfill


\bibliographystyle{epj}
\bibliography{biblio}

\end{document}